\newcommand{\version}{January 12, 2016}
\documentclass[a4paper,oneside,11pt,pdftex]{article}
\pdfoutput=1
\usepackage[bindingoffset=0.0cm,textheight=22.5cm,hdivide={*,16.0cm,*}, vdivide={*,22cm,*}]{geometry}
\usepackage{amsmath,amsfonts,amssymb}
\usepackage[T1]{fontenc}
\usepackage[utf8]{inputenc}
\usepackage{lmodern}
\usepackage[pdftex]{graphicx}
\usepackage[backend=bibtex,bibencoding=ascii,sorting=none,sortcites=true,citestyle=numeric-comp,bibstyle=custom,maxnames=6,minnames=3,firstinits=true]{biblatex}
\usepackage[T1]{fontenc}
\usepackage[utf8]{inputenc}
\usepackage{lmodern}
\usepackage[font=small]{caption}
\usepackage[pdftex,hyperref,svgnames]{xcolor}
\usepackage[pdftex,bookmarksnumbered=true,breaklinks=true,%
colorlinks=true,linktocpage=true,linkcolor=MediumBlue,citecolor=ForestGreen,urlcolor=DarkRed]{hyperref}
\makeatletter
\AtBeginDocument{
  \hypersetup{
    pdfsubject = {\Abstract},
    pdfkeywords = {\keywords},
    pdftitle = {\@title},
    pdfauthor = {\@author}
  }
}
\makeatother
\newcommand{\acknowledgments}{\subsection*{Acknowledgments}}
\makeatletter

 \@addtoreset{equation}{section}
 \makeatother
\IfFileExists{dsfont.sty}
	{\usepackage{dsfont}
         \let\mathbb=\mathds
         \newcommand{\id}{\mathds{1}}}
	{\typeout{Package dsfont.sty was not found, using alternative macros.}
         \let\mathds=\mathbb
         \newcommand{\id}{\mbox{1 \kern-.59em \textrm{l}}}}

\usepackage{mathtools}
\usepackage{slashed}
\usepackage{tensind}
\tensordelimiter{?}

\newcommand{\moyal}{Groenewold-Moyal}
\newcommand{\uim}{UV/IR mixing}
\newcommand{\nc}{non-commu\-ta\-tive}

\newcommand{\eqnref}[1]{Eqn. (\ref{#1})}
\newcommand{\figref}[1]{Figure \ref{#1}}

\newcommand{\secref}[1]{Section \ref{#1}}

\newcommand{\starco}[2]{\left[ #1\stackrel{\star}{,}#2\right] }
\newcommand{\staraco}[2]{\left\{ #1\stackrel{\star}{,}#2\right\} }
\newcommand{\pa}{\partial}

\newcommand{\ri}{\mathrm{i}}

\renewcommand{\k}{\tilde{k}}
\newcommand{\p}{\tilde{p}}

\renewcommand{\a}{\alpha}
\renewcommand{\b}{\beta}
\newcommand{\g}{\gamma}
\renewcommand{\d}{\delta}

\renewcommand{\th}{\theta}

\renewcommand{\l}{\lambda}
\newcommand{\m}{\mu}
\newcommand{\n}{\nu}

\renewcommand{\r}{\rho}
\newcommand{\s}{\sigma}
\renewcommand{\t}{\tau}

\newcommand{\W}{\Omega}

\newcommand{\inv}[1]{\frac{1}{#1}}
\newcommand{\tinv}[1]{\tfrac{1}{#1}}

\newcommand{\nn}{\nonumber}

\newcommand{\ig}{\mathrm{i}g}

\DeclareMathOperator{\tr}{tr} 
\DeclareMathOperator{\Tr}{Tr}

\newcommand{\txt}[1]{\textrm{#1}}

\newcommand{\coleq}{\vcentcolon=}

\newcommand{\Title}{Aspects of perturbative quantum field theory on non-commutative spaces}

\newcommand{\Abstract}{%
In this contribution to the proceedings of the Corfu Summer Institute 2015, I give an overview over quantum field theories on non-commutative Moyal space and renormalization.
In particular, I review the new features and challenges one faces when constructing various scalar, fermionic and gauge field theories on Moyal space,
and especially how the {\uim} problem was solved for certain models.
Finally, I outline more recent progress in constructing a renormalizable gauge field model on {\nc} space, and how one might attempt to prove renormalizability of such a model using a generalized renormalization scheme adapted to the {\nc} (and hence non-local) setting.
}
\newcommand{\keywords}{non-commutative geometry, quantum field theory, renormalization}

\title{\texorpdfstring{\begin{flushright}
        {\small LA-UR-16-20147}
       \end{flushright}\vspace{2em}}{}%
       \Title}
\author{Daniel N. Blaschke\texorpdfstring{\footnote{Presented at the Corfu Summer Institute 2015 ``School and Workshops on Elementary Particle Physics and Gravity'', 1-27 September 2015, Corfu, Greece.}}{}}
\date{\version\\[1em]
\normalsize
Los Alamos National Laboratory\\Los Alamos, NM, 87545, USA
\\[0.5cm]
\ttfamily{E-mail: dblaschke@lanl.gov}}

\graphicspath{{./}{./figures/}}

\newcommand{\pos}[1]{to appear in \href{http://pos.sissa.it/cgi-bin/reader/conf.cgi?confid=263}{\textit{PoS} (CORFU2015)}}

\begin{document}

\maketitle

\abstract{\Abstract}

\newpage
\tableofcontents

\section{Introduction and motivation}

The main motivation to study quantum aspects of space-time is the fact that the classical concept of space and time must break down at Planck scale (or possibly even larger) distances.
During the past century, two very successful theories have been developed which describe the fundamental forces in nature to high experimental precision:
on microscopic scales Quantum Field Theory (QFT) and based thereon the Standard Model of Particle Physics, and on macroscopic scales Einstein's Theory of General Relativity (GR) describing gravity in terms of space-time curvature.
Unfortunately, these two theories are incompatible with each other for the following reason:
GR is described by the Einstein equations
\begin{align}
R_{\m\n}-\inv{2}Rg_{\m\n}&=\langle \hat{T}_{\m\n}\rangle
\,,
\end{align}
whose left hand side is governed by purely geometrical objects, i.e. metric and curvature.
On the right hand side, the ``source'' of space-time curvature, the energy-momentum tensor $\hat{T}_{\m\n}$ of matter, must be a quantum mechanical object, as matter is described by quantum field theory.
Consistency can be achieved by taking the classical approximation of the energy-momentum tensor (denoted by $\langle\ldots\rangle$ above). Since, gravity is in general a very weak force relevant only at macroscopic distances, such an approximation is justified in most cases.
However, this situation is unsatisfactory not only from a mathematical point of view, but also considering that there are indeed situations where quantum effects will play a major role.
The most prominent examples are black holes, whose space-time singularities can be expected to be regularized by such quantum effects.
Although the exact length scale at which quantum geometric effects start to become important is not yet known, one can argue that it should be at least of order of the Planck length $\lambda_p=\sqrt{{G\hbar}/{c^3}}\simeq10^{-33}\textrm{cm}$, as the energy required to resolve that length scale experimentally would create a black hole of the same size~\cite{Doplicher:1994b}.

Hence, in order to unify all forces of nature in a mathematically consistent way, gravity has to be described by a quantum theory of some sort as well.
Here, I consider the non-commutative geometry approach and concentrate on the construction of quantum field theories on flat non-commutative spaces, in particular on aspects of renormalization.

Apart from the question of unification at the Planck scale, there is a further reason to study field theories on non-commutative spaces:
Non-commutative geometry can appear effectively in certain limits of other physical problems.
In particular, the Quantum Hall Effect can be described by effective non-commutative field theories.
For details I refer the interested reader to Refs.~\cite{Susskind:2001fb,Polychronakos:2007,Pasquier:2007nda}.

The present work is organized as follows:
In \secref{sec:deformation} I introduce the notion of a ``quantized space'', in particular flat Moyal space.
I then review the construction of scalar, fermion and gauge fields and their respective actions on such a space, emphasizing various features and obstacles of QFTs thereof in \secref{sec:NCQFTs}.
In \secref{sec:renormalization} I introduce some renormalizable field theories on Moyal space and outline the difficulties in constructing a renormalizable gauge field action before, in \secref{sec:techniques}, I present a renormalization scheme which was recently adapted to gauge fields on {\nc} spaces.
I then close the discussion with a brief outlook.

\section{Deformation of spaces}
\label{sec:deformation}

The notion of a ``quantized space'' is closely tied to the idea that some ``minimal length'' of space-time should exist~\cite{Schroedinger:1934}, and was historically motivated by the wish to ``smear out'' point-like interactions of particles in order to regularize ultraviolet divergences~\cite{Snyder:1946} which are typical for quantum field theories.
Due to the success of renormalization procedures, which deal with these divergences, interest in non-commutative geometry was subdued, and finally renewed in the 1990s~\cite{Doplicher:1994b,Madore:1991,Connes:1994,Filk:1996,Connes:1997cr,Douglas:1997,Seiberg:1999,Minwalla:1999px}.
Today, several extensive reviews exist on this field, see for example~\cite{Landi:1997,Szabo:2001,Rivasseau:2007a,Blaschke:2010kw}.
It is also worth mentioning that quantum field theories on non-commutative spaces are closely related to matrix models, where gravity is an emergent force in the semi-classical limit~\cite{Steinacker:2010rh,Blaschke:2011qu,Steinacker:2014fja}.


Since any geometric space may be represented by a commutative $C^*$ algebra, a straightforward generalization to non-commutative spaces is achieved by considering non-commutative $C^*$ algebras~\cite{Connes:1994,Landi:1997}.
Thus, in {\nc} quantum field theories (NCQFTs), the coordinates themselves have to be considered as operators $\hat x^i$ on some Hilbert space $\mathcal{H}$, satisfying an algebra defined by commutation relations of the form
\begin{align}
[\hat x^\m, \hat x^\n] = \ri \th^{\m\n}(\hat x)
\,,\label{eq:commutator}
\end{align}
where $\th^{\m\n}(\hat x)$ might be any function of the generators with $\th^{\m\n} = - \th^{\n\m}$ and satisfying the Jacobi identity.
The commutation relations can be either constant (i.e. the canonical case leading to a Heisenberg-type algebra and uncertainty relation $\Delta x^\mu\Delta x^\nu\geq\frac12 |\theta^{\mu\nu}|\sim(\lambda_p)^2$),  
linear (the Lie-algebra case $[\hat x^\m, \hat x^\n] = \ri \l^{\m\n}_k \hat x^k$ leading to fuzzy spaces \cite{Madore:1990,Madore:1991} and $\kappa$-deformation \cite{Lukierski:1991,Majid:1994,Wohlgenannt:2003}),
or quadratic (i.e. $[\hat x^\m, \hat x^\n] = \ri \l^{\m\n}_k \hat x^k$ corresponding to quantum groups \cite{Reshetikhin:1990,Wess:1994}) in the generators.

Independent of the explicit form of $\th^{\m\n}$, there is an isomorphism mapping of the {\nc} function algebra $\hat{\mathcal A}$ to the commutative one equipped with an additional {\nc} product $\star$, $\{\mathcal A, \star\}$, i.e.:
\begin{align}
\hat{\mathcal{W}}: \mathcal{A} & \to  \hat{\mathcal A} \,,&
x^i & \mapsto  \hat x^i \,,&
x^i x^j & \mapsto \, :\hat x^i \hat x^j: \quad \textrm{ for } i<j
\,, \label{eq:quantization}
\end{align}
where an operator ordering prescription indicated by $:\, :$ has to be defined (see e.g. the review article~\cite{Blaschke:2010kw} and references therein for details).
The according star product is then defined by
\begin{align}
\hat{\mathcal{W}}(f\star g) \coleq \hat{\mathcal{W}}(f) \cdot \hat{\mathcal{W}}(g) = \hat f\cdot \hat g
\,, \label{eq:star_prod_def}
\end{align}
where $f, g \in \mathcal A$, $\hat f, \hat g \in \hat{\mathcal A}$.
If we choose a symmetrically ordered basis, we can use the Weyl-quantization map for $\hat{\mathcal{W}}$:
\begin{align}
&\hat f=\hat{\mathcal{W}}[f]\coleq\int\! d^D\! x\, f(x)\hat\Delta(x)\,,
&&
\hat\Delta(x)=\int\!\frac{d^D\!k}{(2\pi)^D}e^{i k_\mu \hat{x}^\mu}e^{-i k_\mu x^\mu} \,,
\nn\\
&f(x)=\textrm{Tr}\left(\hat{\mathcal{W}}[f]\hat\Delta(x)\right)
\,, &&
\textrm{Tr}\hat{\mathcal{W}}[f]=\int\!{d^D\!x}f(x)
\,, \label{eq:weyl-quant}
\end{align}
where $D$ denotes the dimension of space-time.
Derivations are defined via
$[\hat\partial_\mu,\hat x^\nu]=\delta_\mu^\nu$ leading to the property
$[\hat\partial_\mu,\hat{\mathcal{W}}[f]]=\hat{\mathcal{W}}[\partial_\mu f]$, i.e. the derivative operator $[\hat\partial_\mu,\cdot]$ acing on a Weyl symbol $\hat{\mathcal{W}}[f]$ equals the Weyl symbol of the usual derivative of function $f$.
The exponential $e^{i k_\mu \hat{x}^\mu}$ appearing in $\hat\Delta$ is defined via its Taylor expansion and thus accounts for the symmetrical operator ordering.
Using \eqnref{eq:star_prod_def} we thus get
\begin{align}
\hat{\mathcal{W}}(f \star g)= \int\!\frac{d^D\!k}{(2\pi)^D}\, d^D\!p\,
e^{ik_i\hat x^i}e^{ip_j \hat x^j} \tilde f (k) \tilde g (p)
\,,\label{eq:star2}
\end{align}
where $\tilde f(k) =  \int d^D\!x\, e^{-ik_j x^j} f(x)$ is the Fourier transform of $f$.
Because of the non-commu\-ta\-tivity of the coordinate operators $\hat x^i$, 
we have to apply the Baker-Campbell-Hausdorff (BCH) formula
\begin{align}
e^A e^B = e^{A+B +\inv{2}[A,B] + \inv{12}[[A,B],B] - 
\inv{12} [[A,B],A] + \dots}\,.
\end{align}
For example, considering the canonical case where $\th$ is constant, all higher order terms in the BCH formula vanish, leading to
\begin{align}
\hat{\mathcal{W}}(f \star g)=\inv{(2\pi)^D} \int d^D\!k\,d^D\!p\,
e^{\ri(k+p)\hat x-\frac{\ri}{2} k_\m\th^{\m\n}p_\n} \tilde f (k) \tilde g (p)
\,,\label{eq:moyal-star}
\end{align}
and hence the {\moyal} star product~\cite{Groenewold:1946,Moyal:1949} $f \star g$ may formally be written as
\begin{align}
(f \star g)(x)= e^{\frac{i}{2}\theta^{\mu\nu} \partial^x_\mu\partial^y_\nu} f(x) g(y)\Big|_{x = y}
\neq\left(g \star f\right)(x)
\,.\label{eq:moyal-star-formal}
\end{align}
The generalization to multiple fields is straightforward:
\begin{align}
\left(f_1\star\cdots\star f_m\right)(x)&=\iiint\!\frac{d^D\!k_1}{(2\pi)^D}\cdots\frac{d^D\!k_m}{(2\pi)^D}\,e^{i\sum\limits_{i=1}^mk_i x} 
\tilde{f}_1(k_1)\cdots\tilde{f}_m(k_m)e^{-\frac{i}{2}\sum\limits_{i<j}^mk_i\theta k_j}
\,.
\end{align}
This is the (associative but {\nc}) star product which will be mainly considered in the following.
It is defined for any Schwartz space functions, which is all we will need for field theory (where asymptotic boundary conditions are assumed).
Apart from the above Fourier representation, other forms of the Moyal product can be derived as well~\cite{Rivasseau:2007a}, such as
\begin{align}
(f\star g)(x)&=\int\!\frac{d^D\!k}{(2\pi)^D}\int\!d^D\!z\, f(x+\tinv{2}\th k)g(x+z)e^{\ri k_\m z^\m} \nn\\
&=\inv{\pi^D|\det\th|}\iint\!d^D\!y\,d^D\!z f(x+y)g(x+z)e^{-2\ri y^\m\th^{-1}_{\m\n}z^\n}
\,, \label{eq:WM-starprod-2-variations}
\end{align}
where the second line is only true if $\th^{\m\n}$ is invertible.
The last version of the star product enables us to compute the star product of two Dirac delta functions:
\begin{align}
\d^D(x)\star \d^D(x)&=\inv{\pi^D|\det\th|}
\,,
\end{align}
i.e. the star product of two point sources becomes infinitely non-local. This means that very high energy processes can have important long-distance consequences.

Since integrations of star products correspond to traces on the operator side, cf. \eqref{eq:weyl-quant}, invariance under cyclic permutations is inherited, i.e.
\begin{align}
 \int\! d^D\!x\, \left(f \star g\star h\right)(x)=\int\! d^D\!x\, \left(h\star f \star g\right)(x)
 \,. \label{eq:cyclic-inv}
\end{align}
Finally, we will also need functional variations in order to define quantum field theories:
\begin{align}
 \frac{\delta\  }{\delta f_1(y)}\int\! d^D\!x\left( f_1\star f_2\star\cdots\star f_m\right)(x)=\left( f_2\star\cdots\star f_m\right)(y)
 \,.
\end{align}

\section{Quantum field theory on non-commutative spaces}
\label{sec:NCQFTs}
\subsection{Scalars}

Let us illustrate some basic properties of quantum field theories on Euclidean Moyal space by means of a scalar $\phi^4$ theory\footnote{
We will not discuss the transition from Euclidean to Minkowskian signature (or vice versa) and time-ordering.
For references on this open and very interesting question, see e.g. \cite{Bahns:2009,Grosse:2011es} and references therein.}.
The straightforward (or naive as will see shortly) generalization is achieved by replacing all fields $\phi$ by operators $\hat\phi=\hat{\mathcal{W}}[\phi]$ and subsequently employing the Weyl quantization introduced above:
\begin{align}
S &=\textrm{Tr}\left(\frac1{2}[\hat\partial_\mu,\hat{\mathcal{W}}[\phi]]^2+\frac{m^2}{2}\hat{\mathcal{W}}[\phi]^2+\frac{\lambda}{4!}\hat{\mathcal{W}}[\phi]^4\right) \nn\\
&= \int\!d^4x \left( \frac1{2}\partial_\mu \phi \star \partial^\mu \phi + \frac{m^2}{2} \phi \star \phi + \frac\lambda{4!} \phi\star\phi\star\phi\star\phi \right)
\,. \label{eq:scalar-action-naive}
\end{align}
Cyclic invariance of the star product under the integral implies that $\int\! d^4\!x \left(f \star g\right)\!(x)=\int\! d^4\!x\, f(x) g(x)$, see \eqref{eq:cyclic-inv}.
Thus, bilinears of any QFT action are unaffected by the star product, and so are the propagators iff Euclidean spaces are considered.
Vertices on the other hand pick up additional phase factors.
These phases act as regulators in Feynman diagrams, leading to finite results where in commutative space ultraviolet divergences would have been expected.
For example, the self energy of our scalar field at one loop order has two contributions, $\int\!d^4k (k^2+m^2)^{-1}$ exhibiting the usual quadratic ultraviolet divergence known from its commutative counterpart, and
\begin{align}
 \frac14 \int d^4k \frac{e^{i k_\m\th^{\m\n} p_\n}}{k^2+m^2}=\sqrt{\frac{m^2}{\p^2}}K_1\!\!\left(\sqrt{m^2\p^2}\right)
 \,, \qquad \p^\m\coleq\th^{\m\n}p_\n
 \,,
\end{align}
where $K_1(z)$ is the modified Bessel function.
The result is finite, but for small $\th$ or external momentum $p$ it behaves as $\p^{-2}+\txt{const.}m^2\ln(m^2\p^2)$ exhibiting a quadratic and a logarithmic infrared divergence.
The reason is quite simple:
When $\p^\m$ tends to zero, the regulating effect of the phase factor is lost and the quadratic ultraviolet divergence has to reappear in the result, manifesting itself as an infrared divergence in the regulator,
and thus giving this effect its name: {\uim}~\cite{Minwalla:1999px,Matusis:2000jf,Micu:2000}.
The interpretation of this mixing of scales, however, is more subtle.
E.g. it can be seen as being due to the coupling of fields to (emergent) gravity~\cite{Steinacker:2010rh}.
In any case, these new types of IR divergences cannot be regularized using a mass and present a true obstacle to renormalization:
Imagine a higher loop Feynman graph which includes multiple self-energy insertions.
In such a diagram arbitrary powers of $1/\p^2$ can appear and hence the outer loop integral over $p$ fails at $p=0$.
In order to render the theory renormalizable, additional relevant operators must be added to the action\footnote{
Another path one might take, is to impose supersymmetry~\cite{Chepelev:2000} as is done in the matrix model approach~\cite{Steinacker:2010rh}.
We will, however, not discuss supersymmetry here.}, as we will outline in \secref{sec:renormalization}.

\subsection{Gauge fields}

When generalizing gauge fields to the {\nc} setting, one has to consider always enveloping algebras such as $U(N)$, $O(N)$, etc. since $SU(N)$, $SO(N)$, etc, do not close~\cite{Armoni:2000xr,Armoni:2001}.
This can be easily seen from the star commutator of two Lie algebra valued functions $\a=\a^a T^a$, $\b=\b^a T^a$ with generators $T^a$,
\begin{align}
 [\alpha\stackrel{\star},\beta] = \frac1{2}\{\alpha^a\stackrel{\star},\beta^b\} [T^a,T^b] + \frac1{2}[\alpha^a\stackrel{\star},\beta^b] \{T^a,T^b\}
 \,,
\end{align}
which features a second term, that is proportional to the anti-commutator of the generators and which vanishes only in the commutative limit where $\lim\limits_{\th\to0}[\alpha^a\stackrel{\star},\beta^b]=0$.
For now, we take the $T^a$ to be $U(N)$ generators.
As in the scalar case, the strategy is to replace all fields with operators and employ the Weyl quantization prescription leading to
\begin{align}
&S=\!\frac1{4}\Tr\big([\hat\partial_\mu,\!\hat{\mathcal{W}}[A]_\nu]\!-\![\hat\partial_\nu,\!\hat{\mathcal{W}}[A]_\mu]\!-\!\ig[\hat{\mathcal{W}}[A]_\mu,\!\hat{\mathcal{W}}[A]_\nu]\big)^{\!2} \nn\\
&\phantom{S} =\frac1{4}\int\!d^Dx\,\tr_N\left(F_{\mu\nu} \star F^{\mu\nu}\right)\,, \nn\\
&F_{\mu\nu}=\partial_\mu A_\nu-\partial_\nu A_\mu-\ig[A_\mu\stackrel{\star},A_\nu]
\,, \label{eq:action-GFT}
\end{align}
where the remaining trace is over $U(N)$.
This action is invariant under the infinitesimal gauge transformations
\begin{align}
\delta_\alpha A_\mu &=D_\mu \alpha =\partial_\mu \alpha -\ig[A_\mu \stackrel{\star},\alpha] \,, &
\delta_\alpha F_{\mu\nu} &=-\ig[F_{\mu\nu} \stackrel{\star},\alpha]
\,.
\end{align}
Due to the non-commuting nature of the star product, even the $U(1)$ case leads to a non-Abelian structure, i.e. the star-commutators in the field strength and the gauge transformations are always present,
and the field strength is only gauge covariant (not invariant).
Likewise, the Lagrangian is not gauge invariant, but transforms covariantly:
Only the action is invariant, since it is the integral which renders expressions invariant under cyclic permutations.

Because of the relations
\begin{align}
\starco{f(x)}{g(x)}&=2\ri \sin\left(\tinv{2}\th^{\m\n}{\pa_\m^x}{\pa_\n^y}\right)f(x)g(y)\Big|_{x = y}\,, \nn\\
\staraco{f(x)}{g(x)}&=2\cos\left(\tinv{2}\th^{\m\n}{\pa_\m^x}{\pa_\n^y}\right)f(x)g(y)\Big|_{x = y}
\,, \label{eq:star-properties-1}
\end{align}
sine and cosine functions play the role of (anti-)symmetric structure ``constants'' of the star-non-Abelian algebra which are non-zero even in the $U_\star(1)$ case.
These non-Abelian properties also lead to features such as asymptotic freedom of the coupling~\cite{Armoni:2000xr,Martin:2000bk,Armoni:2001},
which are typical for non-Abelian gauge theories on commutative space such as QCD,
but not what we would expect from $U(1)$ gauge fields or photons.
Additionally, ghosts do not decouple from $U(1)$ gauge fields.
For example, the BRS invariant action in Landau gauge fixing reads
\begin{align}
S&=\frac1{4}\int\!d^4x\,\tr_N\big(F_{\mu\nu}\star F^{\mu\nu}
+b\star\partial^\mu A_\mu -\bar{c}\star\partial^\mu D_\mu c\big)
\,,
\end{align}
and is invariant under the nilpotent supersymmetric BRS transformations
\begin{align}
s A_\mu&=D_\mu c=\partial_\mu c-\ig[A_\mu\stackrel{\star},c] \,,
& sc&=\ig c\star c \,, 
\nn\\
s\bar c&=b \,, 
& sb&=0 \,, \nn\\
s^2\varphi&=0\,, \quad \forall\varphi
\,,
\end{align}
where $\bar c$, $c$ are the (anti-)ghost fields and $b$ is the Lagrange multiplier implementing the Landau gauge fixing.

\begin{figure}[ht]
 \centering
 \includegraphics[width=0.3\textwidth]{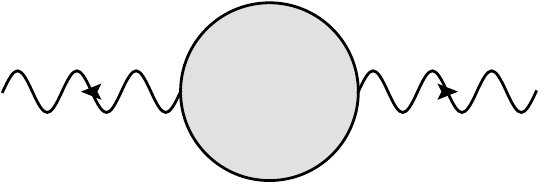}\hspace*{2cm}%
 \includegraphics[width=0.3\textwidth]{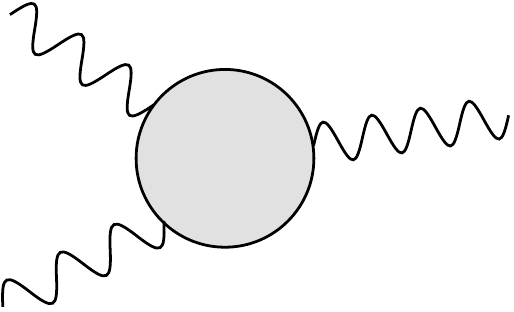}
 \caption{The one-loop corrections to the gauge field two-point and three-point function exhibit quadratic and linear infrared divergences, respectively.}
 \label{fig:gaugeloops}
\end{figure}

As in the scalar case above, gauge theories on Moyal space suffer from {\uim} --- see~\cite{Martin:1999aq,Hayakawa:1999,Martin:2000,Ruiz:2000} for some early papers on {\nc} gauge field theories (NCGFTs).
In particular, at one-loop order (see \figref{fig:gaugeloops}) the vacuum polarization exhibits a quadratic IR divergence
\begin{align}
 \Pi^{\textrm{IR}}_{\mu\nu}(p)\propto\frac{\tilde p_\mu\tilde p_\nu}{(\tilde p^2)^2}
 \,,
\end{align}
even though gauge invariance restricts the UV divergence to being only logarithmic.
Likewise, the one-loop correction to the 3-point function exhibits a linear infrared divergence
\begin{align}
\Gamma^{3A,\textrm{IR}}_{\mu\nu\rho}(p_1,p_2,p_3)
&\propto\cos\left(\frac{p_1\p_2}{2}\right)\sum\limits_{i=1,2,3}\frac{\tilde p_{i,\mu}\tilde p_{i,\nu}\tilde p_{i,\rho}}{(\tilde p_i^2)^2}
\,.
\end{align}

\noindent
The reason for this is that these two IR divergent terms are consistent with the Slavnov-Taylor identities, which in the $U(1)$ case read~\cite{Blaschke:2013gha}
\begin{align}
 \partial^z_\mu\frac{\delta^2 \Gamma^{(0)}}{\delta A_\nu(y)\delta A_\mu(z)}\Big|_{\Phi=0}&=0
 \,,\nn\\
  \partial^z_\mu\frac{\delta^3\Gamma^{(0)}}{\delta A_\sigma(x)\delta A_\nu(y)\delta A_\mu(z)}&=\ig \left[{\frac{\delta\Gamma^{(0)}}{\delta A_\sigma(x)\delta A_\nu(y)}\stackrel{\star}{,}\delta(y-z)}\right]
 +(\sigma,x)\leftrightarrow(\nu,y)
 \,,
\end{align}
while no quadratically or linearly UV divergent terms of the right dimension are allowed by them.
A curious feature of NC gauge theories is that these infrared divergences appear only for external $U(1)$ legs in a more general $U(N)$ theory, at least to one-loop order~\cite{Armoni:2000xr}.
Furthermore, these IR divergences are independent from the chosen gauge fixing~\cite{Ruiz:2000,Blaschke:2005b}.

\subsection{Fermions}

Following the same strategy as in the previous subsections, Weyl quantization leads to the following action for charged fermions (in the fundamental representation)~\cite{GraciaBondia:2000pz,Moreno:2000kt}:
\begin{align}
 S&=\int\! d^4x \, \ri \bar\psi\star \gamma^\mu \left( \partial_\mu\psi-\ig A_\mu\star\psi\right)
 \,,
\end{align}
where the $\g$-matrices fulfill the usual Clifford algebra $\{\gamma^\mu,\gamma^\nu\}=2\eta^{\mu\nu}$.
This action is invariant under the gauge transformations
\begin{align}
 \delta_\lambda\psi &=\ig\lambda\star\psi\,, &
  \delta_\lambda\bar\psi &=-\ig\bar\psi\star\lambda \,, &
  \delta_\lambda A_\mu &=\partial_\mu\lambda-\ig\left[A_\mu\stackrel{\star}{,}\lambda\right]
  \,.
\end{align}

In contrast to commutative space, however, Moyal space additionally allows couplings between gauge fields and neutral fermions (in the adjoint representation).
This peculiar feature of non-commutative space means that neutrinos can couple to gauge fields in {\nc} space~\cite{Hayakawa:1999,Chaichian:2001mu,Schupp:2002up}, thus providing a means to estimate lower bounds on the NC scale~\cite{Schupp:2002up,Horvat:2011qn}.
The according action reads
\begin{align}
 S&=\int\! d^4x \, \ri \bar\psi\star \gamma^\mu D_\mu\psi
=\int\! d^4x \, \ri \bar\psi\star \gamma^\mu \left( \partial_\mu\psi -\ig \left[A_\mu\stackrel{\star}{,}\psi\right]\right)
\,,
\end{align}
and is invariant under the gauge transformations
\begin{align}
 \delta_\lambda\psi &=-\ig\left[\lambda\stackrel{\star}{,}\psi\right]\,, &
  \delta_\lambda\bar \psi &=-\ig\left[\bar \psi\stackrel{\star}{,}\lambda\right]
  \,.
\end{align}
The coupling is linear in the $\th$-matrix defining the NC scale and thus vanishes in the commutative limit, as does the related current:
\begin{align}
 J^\mu& \equiv
 \frac{\delta S}{\delta A_\mu}=-g\gamma^\mu_{\alpha\beta}\left\{\psi_\beta\stackrel{\star}{,}\bar\psi_\alpha\right\}
\propto \theta
\,,
\end{align}
where $\a,\b$ are spinor indices.

New features like this new coupling are of particular interest when search for signals beyond the Standard model, and it hence makes sense to study small deviations therefrom by constructing a NC version of the Standard Model and expanding all quantities in powers of $\th$.
Such an expansion is known as the Seiberg-Witten map~\cite{Seiberg:1999,Jurco:2001a}.
For details on the SW map expanded NC Standard Model, see e.g. ~\cite{Wohlgenannt:2001,Melic:2005am,Trampetic:2007hx} as well as P. Aschieri's contribution to the present proceedings~\cite{Aschieri:2016PoS}.
Since the expansion in $\th$ has to be truncated at some point, crucial features such as {\uim} are lost.
An alternative SW map which keeps all orders in $\th$ while expanding in NC deviations of the fields from their commutative counterparts has therefore been studied as well~\cite{Schupp:2008,Horvat:2011iv,Horvat:2015aca}, see also J. You's contribution to these proceedings~\cite{You:2016PoS}.

\subsection{Other aspects}

\subsubsection*{The stress tensor}

All actions introduced above lead to NC generalizations of the energy-momentum (or stress) tensor.
For example, the stress tensor derived from the gauge field action \eqref{eq:action-GFT} reads~\cite{AbouZeid:2001up,Grimstrup:2002xs,Das:2002jd}
\begin{align}
 T^{\mu\nu}&= \frac12\Big(\!\left\{F^{\mu\rho}\stackrel{\star}{,}?F^{\nu}_{\rho}?\right\}-\frac12
 \delta^{\mu \nu}
 F_{\rho\sigma}\star F^{\rho\sigma}\Big)
 \,,
\end{align}
which is neither conserved nor gauge invariant.
Instead, it transforms covariantly under gauge transformations and is covariantly conserved:
\begin{align}
 \delta_\lambda T^{\mu\nu}&=-\ig\left[T^{\mu\nu}\stackrel{\star}{,}\lambda\right]
 \,,&
 D_\mu T^{\mu\nu} &= 0
 \,.
\end{align}
The reason is that $T^{\m\n}$ (like the Lagrangian) is a local, non-integrated quantity.
But in Moyal space, the integral plays the role of a trace, and hence only integrated quantities feature cyclic invariance which would be required here for the desired properties.
For the same reason, the stress tensor is in general not unique: cyclic permutations in the star product are allowed as well~\cite{Balasin:2015hna}.

It is, however, possible to construct gauge invariant observables out of gauge covariant quantities using Wilson lines~\cite{Gross:2000ba,AbouZeid:2001up}:
\begin{align}
 \tilde{T}^{\mu\nu}(y)& \equiv
 \int\!\frac{d^4k\, d^4x}{(2\pi)^4} \, e ^{i k(y-x)}\star W(k,x)\star T^{\mu\nu}(x)\,,\nn\\
 W(k,x)&={\cal P}_\star\exp\left(\int_0^1\!d\sigma\,A_\mu(x+\sigma\theta k) \, \theta^{\mu\nu}k_\nu\right)
 \,,
\end{align}
where ${\cal P}_\star$ denotes path ordering with respect to the contour parameter $\s$.
Since the Wilson line transforms as $W(k,x)\to U(x)\star W(k,x)\star U(x+\th k)^\dagger$ under a gauge transformation $U(x)$, and since $e^{ikx}$ induces a translation of $U^\dagger$ by $-\th k$, $\tilde{T}^{\mu\nu}$ is found to be gauge invariant.
In the commutative limit, the length of the Wilson line goes to zero and thus $\lim\limits_{\th\to0}\tilde{T}^{\mu\nu}={T}^{\mu\nu}$.

It is then subsequently possible to make a shift in $\tilde{T}^{\mu\nu}$ to render it conserved, but at the price of loosing symmetry in its indices~\cite{AbouZeid:2001up}.
The procedure outlined here works only if the stress tensor ${T}^{\mu\nu}$ is conserved covariantly --- but that is no longer the case as soon as couplings to matter (fermions, scalars, etc.) are considered~\cite{Balasin:2015hna}.
Instead, $D_\mu T^{\mu\nu}$ is proportional to some star commutator (or source) terms which would vanish under an integral.
Such terms are in fact allowed by the {\nc} generalization of Noether's theorem~\cite{Zahn:2003bt}.
It is therefore in general only possible to construct either a gauge invariant but not conserved stress tensor via Wilson lines, or a conserved but not gauge invariant stress tensor via appropriate redefinition, but not both.

\subsubsection*{NC charged ``point'' particles}

As already mentioned in the introduction (in connection with the quantum Hall effect), classical and quantum mechanics on {\nc} space are also of interest as toy models for field theories which are more difficult to handle, such as the case of interactions with gauge fields~\cite{Delduc:2007av,Horvathy:2010wv}.
In this respect, we recall the similar situation of the coupling of matter to Yang-Mills fields on commutative space:
A coarser level of description for these theories was proposed by Wong~\cite{Wong:1970fu} who considered the motion of charged point particles in an external gauge field~\cite{Balachandran:1977ub,Duval:1981js,Kosyakov:1998qi}.
Similar equations may be derived in Moyal space in a ``semi-classical'' approach, where the action of a ``point'' particle coupled to a gauge field reads
\begin{align}
 S[x]&=  -m \int\!d\tau\, \sqrt{\dot{x}^2} - \int d^4y  \,  J^\mu A_\mu \,, \nn\\
 J^\mu(y) & = \int\! d\tau\, q(\tau) \, \dot x^\mu(\tau) \, \delta^4\left(y-x(\tau)\right)
 \,,
\end{align}
where $\t$ parametrizes the particle's trajectory.
Gauge invariance in {\nc} space requires that $D_\m J^\m=0$, and the according equation of motion is~\cite{Balasin:2014dma}
\begin{align}
 m\ddot x^\mu&=q F^{\mu\nu}\dot x_\nu
 \,,
\end{align}
with $F_{\mu \nu}= \partial_\mu A_\nu - \partial_\nu A_\mu - \ig\left[ A_\mu\stackrel{\star}{,}A_\nu\right]$.
This set of equations (i.e. the e.o.m. together with the covariant conservation of the current) constitute the NC analog of Wong's equations for non-Abelian point particles, see~\cite{Balasin:2014dma} and references therein.

\section{Renormalization of non-commutative QFTs}
\label{sec:renormalization}

So far, only very few QFTs on (Euclidean) Moyal space have been found that are renormalizable to all orders in perturbation theory, and none of them involve gauge fields.
On recent progress on the passing to Minkowski space, we refer to~\cite{Grosse:2011es}, where Wick rotation was generalized to the degenerate Moyal case.

\subsubsection*{The Grosse-Wulkenhaar model}

Historically, the first renormalizable NCQFT was the scalar Grosse-Wulkenhaar (GW) model \cite{Grosse:2003,Grosse:2004b,Rivasseau:2005a} which more recently was found to be non-perturbatively solvable~\cite{Grosse:2012uv,Grosse:2015fka} and whose action reads
\begin{align}
  S&=\int\!d^4x\left(\tfrac{1}{2}\partial_\mu\phi\star\partial^\mu\phi +\tfrac{m^2}{2}\phi^{\star2}
 +2\Omega^2(\tilde x_\mu\phi)\!\star\!(\tilde x^\mu\phi) +\tfrac{\lambda}{4!}\phi^{\star4}\!\right)
\,, &
 \tilde{x}_\mu&\coleq(\theta^{-1})_{\mu\nu}x^\nu
 \,.
\end{align}
Compared to the naive scalar action of \eqnref{eq:scalar-action-naive}, an additional harmonic oscillator-like term is present, which ultimately cures the infamous {\uim} problem.
The propagator is the inverse of the operator $\left(-\square+4\Omega^2\tilde{x}^2+m^2\right)$ and is known as the Mehler kernel, which features a damping behavior for high momenta (UV) as well as for low momenta (IR).
An important feature of the GW action is that it is "Langmann-Szabo" invariant~\cite{Langmann:2002}, i.e. up to rescalings by $\W$, it is form invariant under Fourier transformations:
\begin{align}
 S[\phi;m,\lambda,\Omega]\mapsto\Omega^2S[\phi;\frac{m}{\Omega},\frac{\lambda}{\Omega^2},\frac1{\Omega}]
 \,,
\end{align}
although the kinetic term and the oscillator type $\tilde x^2$-term exchange their roles in Fourier space.
Another way to exhibit the duality between the latter two terms is to use the properties that star-commutators with $x$ generate derivations while star-anticommutators with $x$ become pointwise multiplications with $x$.
Then those two terms can be written as
\begin{align}
 \tfrac{1}{2}\partial_\mu\phi\star\partial^\mu\phi
 +2\Omega^2(\tilde x_\mu\phi)\!\star\!(\tilde x^\mu\phi) &= -\tfrac{1}{2}\starco{\tilde x_\m}{\phi}\star \starco{\tilde x^\m}{\phi} +\tfrac{\W^2}{2}\staraco{\tilde x_\m}{\phi}\star \staraco{\tilde x^\m}{\phi}
 \,.
\end{align}

The GW model is not only renormalizable, but also free of the Landau ghost problem~\cite{Grosse:2004a,Rivasseau:2006b}, which constitutes an improvement compared to scalar $\phi^4$ theory on commutative space.
Furthermore, its beta-function vanishes at the self-dual point $\W=1$, whose special role is exhibited by the Langmann-Szabo duality outlined above.
Finally, the oscillator type $\tilde x^2$-term breaks translation invariance, but has been found to have a very nice interpretation in terms of the Ricci scalar of Moyal space~\cite{Buric:2009ss}.

It should also be mentioned, that a generalization of the GW model to the degenerate Moyal plane is possible and renormalizable, if an additional relevant operator is included into the action.
For details, we refer the interested reader to~\cite{Grosse:2008df}.
Furthermore, attempts have been made to generalize this model to Minkowski space~\cite{Fischer:2011wi}.

\subsubsection*{The Gurau et al. model}

\begin{figure}[ht]
 \centering
 \includegraphics[width=0.35\textwidth]{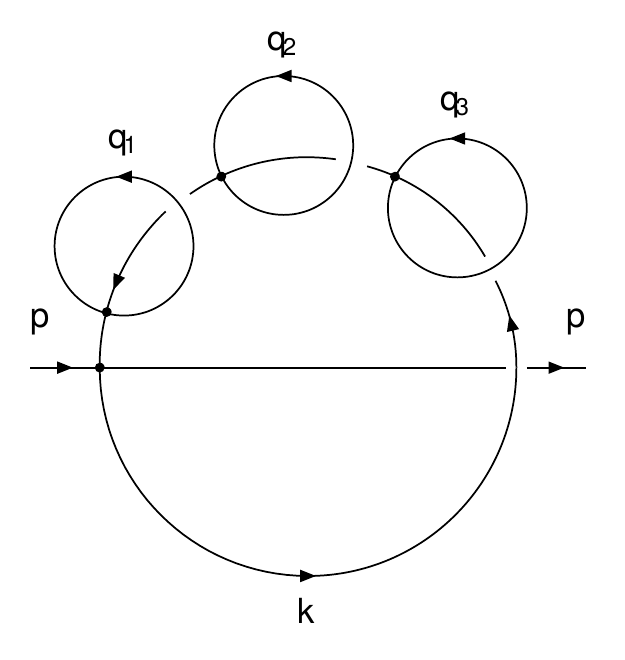}
 \caption{Multiple self-energy insertions into a higher loop graph are rendered finite by the propagator's damping behavior.}
 \label{fig:multiloop}
\end{figure}

An alternative to the GW model is possible for $\phi^4$ theory on Moyal space, which preserves translation invariance.
Its action, which has been proved to be renormalizable by its authors, reads~\cite{Gurau:2009}
\begin{align}
  S=\int\!d^4x\!\left(\!\tfrac{1}{2}\partial_\mu\phi\star\partial^\mu\phi+\tfrac{m^2}{2}\phi^{\star2}
 -\phi(x)\star\frac{a^2}{\widetilde\square_x}\star\phi(x)+\tfrac{\lambda}{4!}\phi^{\star4}\!\right)
 \,, \label{eq:guraumodel}
\end{align}
where $-1/\widetilde\square=-(\th^{\m\r}\th_{\m\s}\pa_{\r}\pa^\s)^{-1}$ becomes $1/\k^2$ in momentum space and $a$ is a dimensionless constant.
The non-local $1/\k^2$-term replaces the GW oscillator term and constitutes a counter term for the one-loop IR divergence of the scalar self-energy.
The propagator features an infrared damping,
\begin{align}
 G(k)&=\frac1{k^2+m^2+\frac{a^2}{{\k}^2}} \,,&
 \lim\limits_{k \to 0}G(k)&=0
 \,,
\end{align}
which is key to the renormalizability of the model and is responsible for rendering insertions of infrared divergent self-energies into higher order loops finite, e.g.
\begin{align}
&\quad\Pi^{n \text{ np-ins.}}(p)\approx\lambda^2\int\!d^4k
\, \frac{e^{i k\p}}{\left({\k}^2\right)^n\left[k^2+m^2+\frac{a^2}{{\k}^2}\right]^{n+1}}
\,.
\end{align}
If $a=0$, this $n+1$ loop graph, which is depicted in \figref{fig:multiloop}, is IR divergent for $n\geq2$.
However, for finite $a$, the integrand remains finite because one has
\begin{align}
 \lim\limits_{k\to0}\frac1{({\k}^2)^n\big[\frac{a^2}{{\k}^2}\big]^{n+1}} =\lim\limits_{k\to0}\frac{{\k}^2}{(a^2)^{n+1}}
 \,.
\end{align}

\subsubsection*{The NC Gross-Neveu model}

The only currently known renormalizable model on Moyal space which includes fermions, is a NC extension of the Gross-Neveu model\footnote{
The Gross-Neveu model is the non-Abelian generalization of the Thirring model.
In Moyal space, however, one always has non-Abelian fields due to the star product.} whose action reads~\cite{VignesTourneret:2007}
\begin{align}
S&=\int d^2x \left[ \bar\psi\left(-i\slashed{\partial}+\Omega\slashed{\tilde{x}}+m\right)\psi+V(\bar\psi,\psi)\right]\,,\nn\\
V(\bar\psi,\psi)&=\l(\bar\psi\star\psi)^{\star2}+\txt{permutations}
\,.
\end{align}
As in the scalar cases, the $\tilde{x}$ dependent term can be interpreted as the coupling to geometry of NC space, although here it is the coupling to torsion in Moyal space~\cite{Buric:2015vja}.

\subsubsection*{Gauge theories}

Many approaches have been attempted to remedy the renormalizability of gauge fields in Moyal space, but difficulties remain and a renormalization proof is still missing.
Here we just mention a few selected models and refer to the reviews~\cite{Blaschke:2009c,Blaschke:2010kw} for further details.
The most obvious strategy to construct a renormalizable NC gauge field theory, is to use the successful scalar models as starting points.
For example, in coupling the GW model to external gauge fields by replacing $\tilde x_\m\to \tilde X_\mu= \tilde x_\mu + gA_\mu$, an effective gauge field action is induced via a heat kernel expansion~\cite{Wallet:2007c,Grosse:2007}.
The $\tilde X_\mu$ transform covariantly under gauge transformations and are hence called ``covariant coordinates''.
Similar to the GW case, both star-commutators and anticommutators in the $\tilde X_\mu$ appear in the resulting ``induced'' gauge field action, the commutator being related to the field strength via $\starco{\tilde X_\mu}{\tilde X_\nu}=\ig F_{\m\n}-\ri\th^{-1}_{\m\n}$.
Furthermore, starting from the naive NCGFT \eqref{eq:action-GFT} and adding an oscillator type term in a BRS invariant way~\cite{Blaschke:2007b} leads to that same action at the one-loop level~\cite{Blaschke:2009aw}.
The main difficulty of the induced gauge field action lies in the vacuum structure which exhibits tadpoles and is not very well understood~\cite{Wallet:2008a,Martinetti:2013uia}.
Also, calculating the propagator is a non-trivial enterprise.

Another approach that has been attempted, is to implement an IR damping for the gauge field propagator via a translation invariant term inspired by the scalar Gurau model \eqref{eq:guraumodel} introduced above.
This has been done for $U(1)$ gauge fields in~\cite{Blaschke:2009e} and extended to the $U(N)$ case in~\cite{Blaschke:2010ck,Blaschke:2013gha}, using techniques known from the Gribov-Zwanziger action in QCD --- see~\cite{Dudal:2008} and references therein for a review of the latter.
The crucial observation is that the restriction to the first Gribov horizon in QCD alters the gluon propagator in a way that it vanishes in the infrared:
By adding the operator
\begin{align}
\gamma^4 g^2 \int\! d^4x\, f^{abc} A_\mu^b (\mathcal{M}^{-1})^{ad}f^{dec}A^e_\mu
\,, \label{eq:inverseFP}
\end{align}
where $(\mathcal{M}^{-1})^{ad}$ is the inverse Fadeev-Popov operator of QCD and $\g$ is known as the Gribov parameter, the gluon propagator is modified to
\begin{align}
G_{\mu\nu}^{ab}&=\frac{\delta^{ab}}{k^2+\frac{\gamma^2}{k^2}}
\left(\delta_{\mu\nu}-\frac{k_\mu k_\nu}{k^2}\right)
\,.
\end{align}
Subsequently, \eqnref{eq:inverseFP} can be localized by introducing additional fields leading to the Gribov-Zwanziger action.
The infrared behavior of the gluon propagator is exactly what we want in the {\nc} case to damp IR divergences from {\uim}.

More recently, the Gribov problem was studied in the NC setting~\cite{Canfora:2015nsa} leading to the conclusion that indeed Gribov copies exist even in the NC $U(1)$ case.
The obvious question now is, whether both the Gribov and the {\uim} problem might be solved simultaneously via the Gribov-Zwanziger approach.
It is, however, unclear which additional operator to add to the action:
the one proposed in~\cite{Blaschke:2009e,Blaschke:2010ck,Blaschke:2013gha} or a NC analog of the inverse Fadeev-Popov operator like in QCD, or possibly both?

\section{On renormalization techniques for NCQFTs}
\label{sec:techniques}

In order to prove renormalizability of a NC gauge field theory candidate, a renormalization scheme must be employed that is compatible with the underlying {\nc} space as well as with gauge symmetry.

A very successful scheme compatible with Moyal space that was used to prove renormalizability for the scalar models in the preceding sections is called Multiscale Analysis~\cite{Rivasseau:2007a}.
Unfortunately, it breaks gauge invariance making it unfeasible for NC gauge theories.
Several other schemes require locality of the theory and cannot easily be generalized to the NC setting which is inherently non-local.
However, there have been more recent attempts at generalizing the well-known BPHZ (named after their inventors Bogoliubov, Parasiuk, Hepp and Zimmermann) renormalization scheme to Moyal space~\cite{Blaschke:2012ex,Blaschke:2013cba}.
In a nutshell, it consists of a subtraction scheme, a proof of locality of these subtractions, normalization conditions, and overlapping (sub-)divergences are treated using Zimmermanns forest formula.
The most intriguing feature, however, is that divergences are subtracted without requiring regularization, e.g.
\begin{align}
J_{\Gamma}^{\txt{finite}} (\underline{p})
&\equiv  \int\!d^4k \, \left[ 1 - t_{\underline{p}}^{\delta(\Gamma) } \right] \,  I_{\Gamma} (\underline{p},k)
\, ,
\nn\\
\left( t_{\underline{p}} ^{N} I_{\Gamma} \right) (\underline{p} ,k)
&\equiv \sum_{l=0}^{N} \frac{1}{l!} \, p_{i_1}^{\mu_1} \cdots p_{i_l}^{\mu_l}
\, \frac{\partial^l I_{\Gamma}}{\partial p_{i_1}^{\mu _1} \cdots \partial p_{i_l}^{\mu _l} }\big( \underline{p} =\underline{0} ,k \big)
\,,
\end{align}
where $\underline{p}$ collectively denotes all momenta $p_i$.
(In the case of the 2-point function there is only one.)
$\delta(\Gamma)$ is the superficial degree of UV divergence of the graph under consideration.
For example, if the superficial degree of divergence is quadratic, $t^2_p$ is required.

\begin{figure}[ht]
 \centering
 \includegraphics[width=0.4\textwidth]{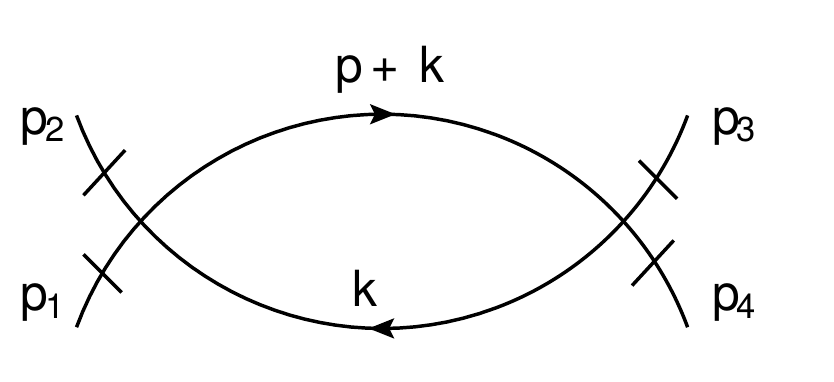}
 \caption{The ``fish'' diagram of $\phi^4$ theory.}
 \label{fig:fish}
\end{figure}

The main problem in applying these subtractions to a NC theory is that the result is \emph{not} finite.
Let us illustrate this point using $\phi^4$ theory's ``fish'' diagram shown in \figref{fig:fish}.
Being superficially a logarithmically divergent one-loop graph, the BPHZ subtraction scheme in commutative space prescribes
\begin{align}
J_{\Gamma}^{\txt{finite}} (p)
 &\equiv  \int\!d^4k \, [1 - t_{p}^{0}  ] \,  I_{\Gamma} (p,k)
  = \int\!d^4k \, [ I_{\Gamma} (p,k) -  I_{\Gamma} (0,k) ]
 \nn\\
 &= \int\!d^4k \, \left(  \frac{1}{[(p+k)^2+m^2][k^2+m^2]} -  \frac{1}{[k^2+m^2]^2}  \right)
 \,.
\end{align}
In Moyal space, however, part of this diagram (the so-called non-planar part) includes regularizing phase factors leading to
\begin{align}
J_{\Gamma}^{\txt{NC}} (p)
 &= \int\!d^4k \, \left(  \frac{A+B\cos(k\tilde p)}{[(p+k)^2+m^2][k^2+m^2]} -  \frac{A+B}{[k^2+m^2]^2}  \right)
 \,,
\end{align}
(where once more $\p^\m=\th^{\m\n} p_\n$), which is not finite.
Not only did the subtraction fail to eliminate the infrared divergence due to {\uim} (parametrized by $B$), it also failed to eliminate the UV divergence of the planar part of the graph (parametrized by $A$).

\begin{figure}[ht]
 \centering
 \includegraphics[width=0.3\textwidth]{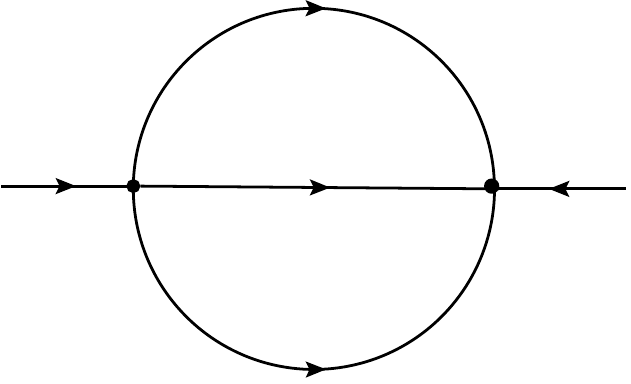}
 \caption{The ``sunrise'' diagram of $\phi^4$ theory features an overlapping divergence.}
 \label{fig:sunrise}
\end{figure}

The remedy which was put forward in~\cite{Blaschke:2012ex,Blaschke:2013cba} suggests to treat $p$ and $\tilde p$ independently in the subtraction scheme, i.e. to modify the subtraction scheme according to
\begin{align}
J_{\Gamma}^{\txt{finite}} (p_i,\tilde{p}_i,k)&=\int\!d^4k \, \left[ 1 - t_{\underline{p}}^{\delta(\Gamma) } \right] \,  I_{\Gamma} (p_i,\tilde{p}_i,k)\,, \nn\\
(t^n_p f)(p_i,\tilde{p}_i)
&\coleq f(0,\tilde{p}_i)+\sum_jp^\mu_j\left(\frac{\partial}{\partial p_j^\mu}f(p_i,\tilde{p}_i)\right)\Big|_{p_i=0}+ \ldots \nn\\
&\quad +\frac1{n!}\sum_{j_1, \dots, j_n}p_{j_1}^{\mu_1}\ldots p_{i_n}^{\mu_n}\left(\frac{\partial}{\partial p_{j_1}^{\mu_1}}\ldots \frac{\partial}{\partial p_{j_n}^{\mu_n}}f(p_i,\tilde{p}_i)\right)\Big|_{p_i=0}
\,.
\end{align}
In our present example this leads to
\begin{align}
J_{\Gamma}^{\txt{NC}} (p)
 &= \int\!d^4k \left(A+B\cos(k\tilde p)\right) \left(  \frac{1}{[(p+k)^2+m^2][k^2+m^2]} -  \frac{1}{[k^2+m^2]^2}  \right)
\end{align}
which is indeed finite.
This modified BPHZ scheme was successfully applied to a scalar $\phi^4$ theory on Moyal space at one-loop level.
Additionally, the ``sunrise graph'', a two-loop graph depicted in \figref{fig:sunrise} which features an overlapping divergence, was studied and successfully treated within this new scheme.

Finally, BPHZ subtractions always involve some ambiguities (finite terms), which in the NC setting lead to additional possible counter terms such as $\phi\widetilde\square^{-1}\phi$ in the scalar case.
This term is precisely the additional one included in the renormalizable Gurau et al. scalar model \eqref{eq:guraumodel}, i.e. that term naturally appears within the modified BPHZ scheme.
The reason we do not find the Grosse-Wulkenhaar oscillator type term is that it breaks translation invariance while for the BPHZ scheme (modified or not) one always enforces translation invariance.
In more general theories, the counter terms generated by the modified BPHZ involve polynomials in $1/\p^2$ (in Fourier space) whose degree is determined by the degree of IR divergence.

The application of this new renormalization scheme to gauge theories was started in~\cite{Blaschke:2013cba} where it was determined that it works fine for the one-loop vacuum polarization.
Since gauge fields are massless, one usually needs a further regularization, but in the Gribov-Zwanziger inspired NC gauge models mentioned in the previous section, the gauge field propagator features an IR damping which renders further regularizations, such as Lowenstein's $s$-trick or dimensional regularization, unnecessary.
Vertex corrections and higher-order loop graphs have not yet been studied.

\section{Conclusion and Outlook}

We have introduced and motivated the concept of QFTs on NC (Moyal) space and given a brief overview over some topics of current research.
While renormalizable scalar and fermionic theories exist on (Euclidean) Moyal space, gauge theories are still an open matter although promising candidates for renormalizable NCGFTs exist.
A proof of renormalizability of gauge theories on Moyal space requires a compatible renormalization scheme, possibly the modified BPHZ scheme outlined in the last section of this paper.

Some open questions that suggest themselves in this work are
\begin{itemize}
\itemsep=3pt
 \item Which one of the two Gribov-Zwanziger inspired operators (or both) should be included in NC gauge field theories in order to render them renormalizable?
 \item A renormalization proof to all orders in perturbation theory is required for NCGFTs, either using the modified BPHZ scheme or developing/generalizing yet another renormalization scheme.
 \item Another possibility is that the translation invariance breaking induced gauge field theory action is renormalizable, but this needs to be proven and the non-trivial vacuum structure needs to be better understood.
 \item Although some progress has been made in recent years, the passing to Minkowski space, especially when NC time is involved, needs to be better understood and worked out for all renormalizable models.
\end{itemize}

\acknowledgments
I would like to thank the organizers for a wonderful and stimulating conference in Corfu.

\printbibliography

\end{document}